\documentclass[nofootinbib,superscriptaddress,twocolumn,showpacs,preprintnumbers,amsmath,amssymb]{revtex4}

\usepackage{graphicx}
\usepackage{grffile}
\usepackage{slashed}
\usepackage{bbm}
\usepackage{footmisc}
\usepackage{multirow}

\usepackage{hyperref}
\usepackage{amssymb}
\usepackage{amsmath}
\usepackage{color}
\usepackage{xspace}

\usepackage{datetime}
\usepackage{color,fancybox}
\newcommand{\GeV}{\ensuremath{\,\mathrm{GeV}}\xspace}
\newcommand{\TeV}{\ensuremath{\,\mathrm{TeV}}\xspace}
\newcommand{\wwz}{WWZ}
\newcommand{\as}{\ensuremath{\alpha_s}}
\newcommand{\aEW}{\ensuremath{\alpha_\text{EW}}}
\newcommand{\nLO}{\nbar \text{LO}\xspace}
\newcommand{\nNLO}{\nbar \text{NLO}\xspace}
\newcommand{\nbar}{\ensuremath{\bar{n}}}
\newcommand{\order}[1]{\mathcal{O}\!\left(#1\right)}

\begin{document}

\title{WZ production beyond NLO for high-\boldmath{$p_T$} observables
}

\preprint{FTUV-12-0920\;\;KA-TP-36-2012\;\;LPN12-102\;\;SFB/CPP-12-70}
\preprint{IPPP/12/70}
\preprint{DCPT/12/140}

\author{Francisco~Campanario}
\email{francisco.campanario@kit.edu}
\affiliation{Institute for Theoretical Physics, KIT, 76128 Karlsruhe, Germany.}
\author{Sebastian~Sapeta}
\email{sebastian.sapeta@durham.ac.uk}
\affiliation{Institute for Particle Physics Phenomenology, Durham
University, South Rd, Durham DH1 3LE, UK.}

\begin{abstract}
  We use the LoopSim and VBFNLO packages to investigate a merged
  sample of partonic events that is accurate at NLO in QCD
  simultaneously for the WZ and WZ+jet production processes.
  In certain regions of phase space %, notably at high transverse momenta,
  such a procedure is expected to account for the dominant part of the
  NNLO QCD corrections to the WZ production process.
  For a number of commonly used experimental observables, we find that
  these corrections are substantial, in the $30-100\%$ range and in
  some cases their inclusion can reduce scale uncertainties by a
  factor of two.
  As in the underlying VBFNLO calculations, we include the leptonic decays
  of the vector bosons and all off-shell and finite-width effects.

\end{abstract}


\maketitle

%------------------------------------------------------------------------------
\section{Introduction}

The study of di-boson production processes at the LHC is important both to test
the Standard Model~(SM) and because they constitute relevant backgrounds to the
beyond standard physics (BSM) searches. 
They are for example sensitive to tri-linear gauge couplings~(TGC) fixed by the
underlying electroweak gauge group in the SM.  Any deviation in the values
observed at experiments would indicate new physics effects. In particular, WZ
production measurements~\cite{ATLAS-WZ-measure,CMS-WZ} are relevant to constrain
the \wwz~ tri-linear coupling and to search for charged heavy bosons from BSM
sector via Jacobian
peaks~\cite{ATLAS-WZ-reso,CMS-WZ-reso,Bagger:1993zf,Englert:2008tn}.

\begin{figure}[h!]
  \begin{minipage}{0.2\columnwidth} 
  \begin{flushright}
  LO 
  \end{flushright}
  \end{minipage}
  \begin{minipage}{0.7\columnwidth}
  \includegraphics[width=0.45\columnwidth]{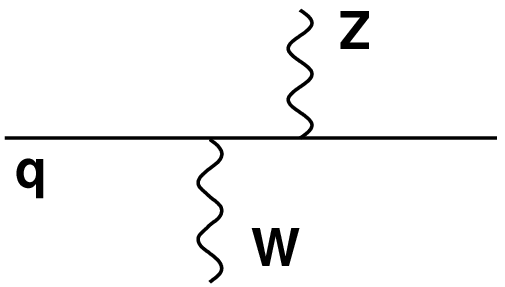}
  \end{minipage}
  \vskip 10pt
  \begin{minipage}{0.2\columnwidth} 
  \begin{flushright}
  NLO 
  \end{flushright}
  \end{minipage}
  \begin{minipage}{0.7\columnwidth}
  \includegraphics[width=0.45\columnwidth]{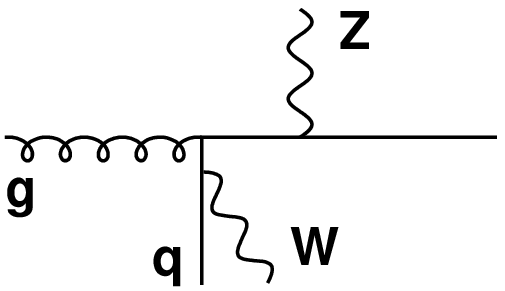}
  \end{minipage}
  \vskip 10pt
  \begin{minipage}{0.2\columnwidth} 
  \begin{flushright}
  NNLO 
  \end{flushright}
  \end{minipage}
  \begin{minipage}{0.7\columnwidth}
  \includegraphics[width=0.45\columnwidth]{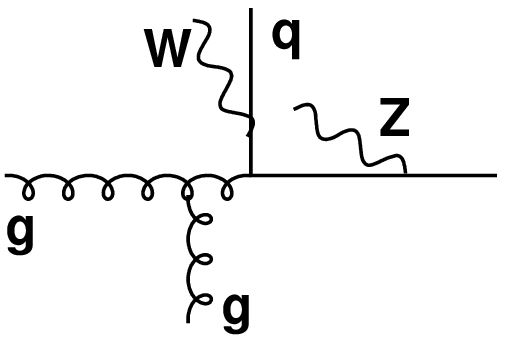}
  \end{minipage}
  \caption{% 
  Example diagram contributing to WZ production at LO, NLO and NNLO.
  }
  \label{fig:diagram}
\end{figure}

To match the experimental accuracy, precise and reliable predictions beyond the
leading order approximation are required. The next-to-leading order~(NLO) QCD
corrections for  WZ production were computed
in~\cite{Ohnemus:1991gb,Frixione:1992pj} and turned out to be sizable.
These large corrections are a well known fact for colorless
production channels. 
At NLO QCD, new topologies and new partonic sub-processes appear,
Fig.~\ref{fig:diagram}, e.g. those with a W or Z emitted from a jet or an
initial-state parton. These new configurations can have partonic
luminosities (e.g. $qg$) that are enhanced  over those present at LO (e.g.
$q\bar q$), which can lead to large corrections in differential
distributions. They also permit soft or collinear
bosons emissions from a quark which results in $\as\aEW^2\ln^2\left(p_{T,j}/m_V\right)$ enhancement
for a number of observables.

%%%%%%%%%%
At the next-to-next-to-leading order~(NNLO) further new sub-processes and
topologies, including those with soft and collinear bosons, start to contribute,
Fig.~\ref{fig:diagram}, which might also lead to large corrections in some
regions of phase space.\footnote{We note that this was the case also for several
observables in di-photon production ~\cite{Catani:2011qz}.} 

Therefore, using only the NLO results to compute the WZ backgrounds may
lead to a misinterpretation of a possible excess observed in data, which could be wrongly
attributed to new physics. Thus, it is  of great interest to try to assess the
genuine NNLO QCD contributions for this process.
The current state of the art is that the virtual two-loop corrections for WZ
production are unknown. 
However, recently, the NLO QCD corrections for WZ+jet were computed
in~\cite{Campanario:2010hp}, including anomalous couplings~\cite{Campanario:2010xn}, and they are available in the VBFNLO
package~\cite{Arnold:2012xn}. 
This result provides the mixed real-virtual and the double real $\order{\as^2}$
contributions to the
WZ production including the NLO corrections to the
$qg$ channel.
This opens the possibility to merge the results for WZ@NLO\footnote{@ in e.g.
WZ@NLO means NLO QCD corrections of WZ.} and WZj@NLO and obtain the part of the
NNLO result for WZ production that is associated with channels absent at LO,
which is the dominant contribution for a range of differential distributions at
high $p_T$.
 
This can be done in a consistent way by employing the LoopSim
method~\cite{Rubin:2010xp} which uses unitarity to cancel infrared and
collinear divergences that arise when one adds
tree level results with different multiplicities. 
The LoopSim prescription is general and it has been found to work
well for Z+jet and Drell-Yan production, especially for 
observables which receive significant corrections at NLO~\cite{Rubin:2010xp}.
The method supplements a set of real and real-virtual diagrams with
approximate contributions from loop diagrams.
The result at a given order is denoted with $\bar n$ to
distinguish it from the exact result. 

In this paper, we compute the \nNLO corrections (according to the
notation introduced above) to WZ production at the LHC
\begin{align}
pp &\rightarrow  W^\pm Z  +
X \rightarrow \ell_1^\pm
\stackrel{\text{\tiny(}-\text{\tiny)}}{\nu_1} \ell_2^+ \ell_2^-  + X\,,
\end{align}
including the leptonic decays and full off-shell and finite width effects. 

We organize this paper as follows: In Section~\ref{sec:caldetails}, we give  
technical details of our computation including a short description of the
LoopSim method. In Section~\ref{sec:numres}, we
present numerical results and the impact of the \nNLO QCD corrections on various
differential distributions. In  Section~\ref{sec:concl}, we summarize our
findings.

%------------------------------------------------------------------------------
\section{Calculational details }
\label{sec:caldetails}

For the calculation of WZ@\nNLO, we used the implementations of WZ@NLO
and WZj@NLO from the VBFNLO package~\cite{Arnold:2012xn}, with the
latter providing the double-real and the real-virtual parts of the NNLO result.
Then the missing two-loop contributions are approximated by LoopSim from the
tree-level and one-loop parts of WZj@NLO.
To allow for communication between VBFNLO and LoopSim, we developed an interface
between the two programs. On one hand, it consists of an extension of VBFNLO
which enables it to write events in the Les Houches (LH) format at NLO and, on
the other hand, of a class which reads the Les Houches events and passes them to
LoopSim. 
Then LoopSim assigns an approximate angular-ordered branching structure to each
LH event, with the help of the Cambridge/Aaachen
(C/A)~\cite{Dokshitzer:1997in,Wobisch:1998wt} jet algorithm. By default, it uses
a radius $R_\text{LS}=1$ and by varying $R_\text{LS}$, one probes one class of
uncertainty of the method.
As we shall see in the next section, except for the very low $p_T$ region, this
uncertainty is significantly smaller than that coming from renormalization and
factorization scale variation.

The error that we make for an observable (A) is a finite constant associated
with the LO topology and it is free of infrared and collinear divergences since
these are, by construction, suitably treated  by LoopSim.
This implies 
\begin{equation}
  \sigma^{(A)}_\text{\nNLO} - \sigma^{(A)}_\text{NNLO} = {\cal
    O}\left(\alpha_s^2\sigma^{(A)}_\text{LO}\right)\,.
\label{eq:acc}
\end{equation}
Thus, differential distributions sensitive to new channels and new kinematically
enhanced configurations resulting in large NLO K-factors, should
be close to the full NNLO result, providing predictions more precise than the
pure NLO corrections to WZ. 

%------------------------------------------------------------------------------
\section{Numerical Results}
\label{sec:numres}

%%%%%%%%%%
\begin{figure}[t]
  \centering
  \includegraphics[width=1.0\columnwidth]{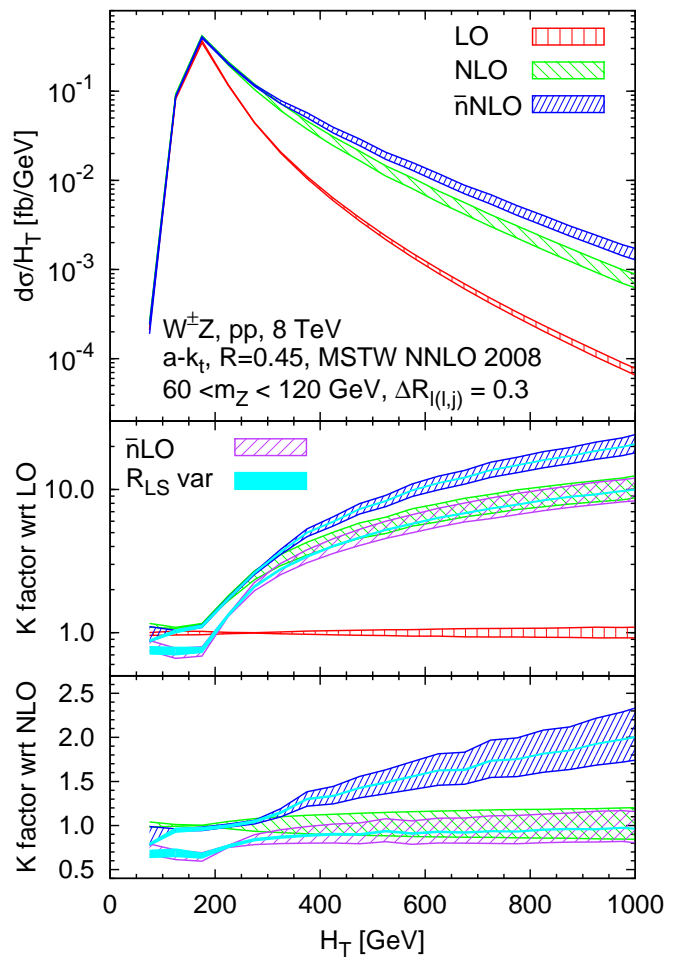}
  \caption{  
  Differential cross sections and K factors for the effective mass observable,
  defined in Eq.~(\ref{eq:HT}), for the LHC at $\sqrt{s}=8\, \text{TeV}$. The
  bands correspond to varying $\mu_F=\mu_R$ by factors 1/2 and 2 around the
  central value from Eq.~(\ref{eq:ren}). The cyan solid bands give the
  uncertainty related to the $R_\text{LS}$ parameter varied between 0.5 and 1.5.
  The distribution is a sum of contributions from two unlike flavor decay
  channels, $ee\mu\nu_\mu$ and $\mu\mu e\nu_e$.
  }
  \label{fig:HT}
\end{figure}

\begin{figure*} [t!]
  \centering
  \includegraphics[width=1.0\columnwidth]{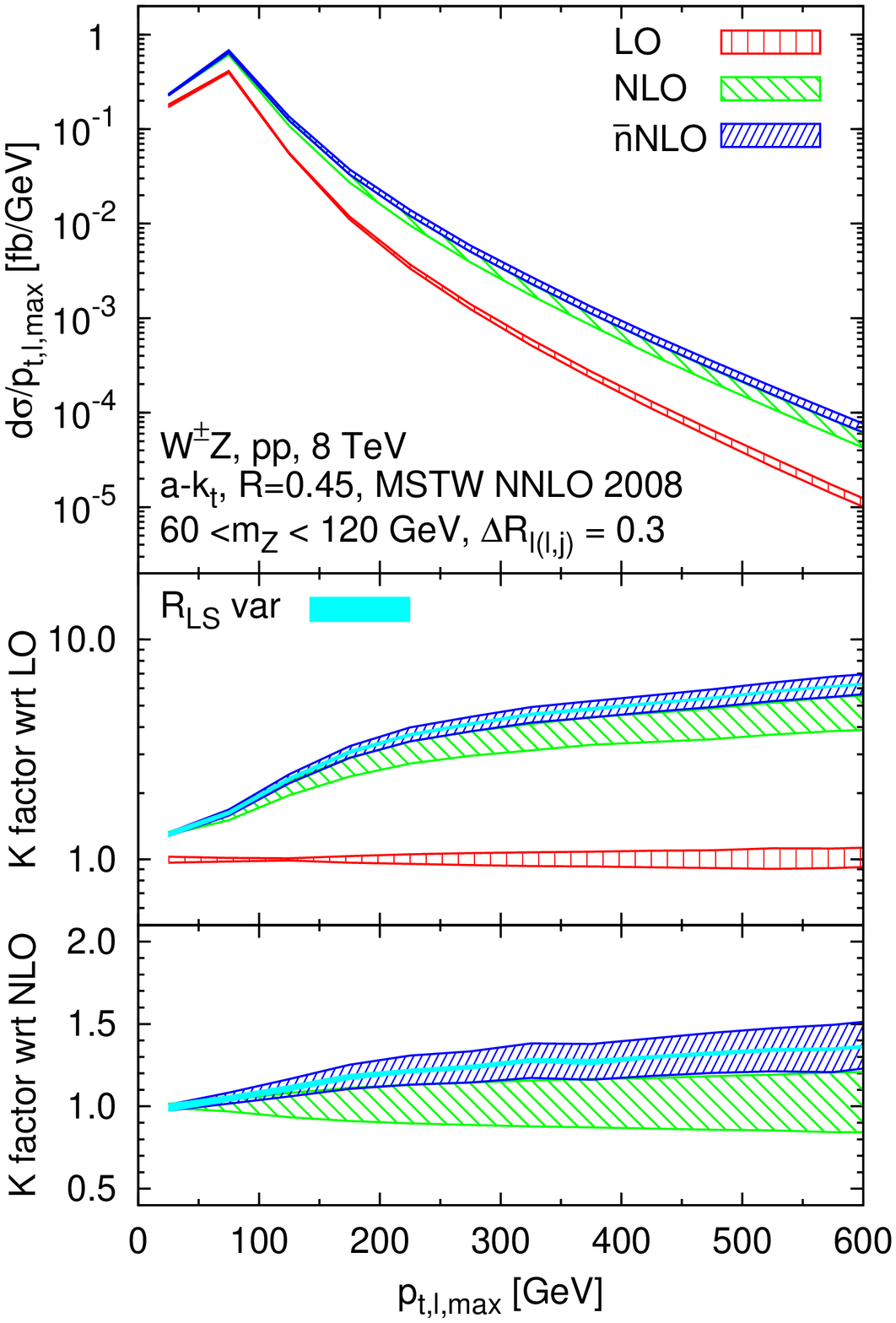}
  \hfill
  \includegraphics[width=1.0\columnwidth]{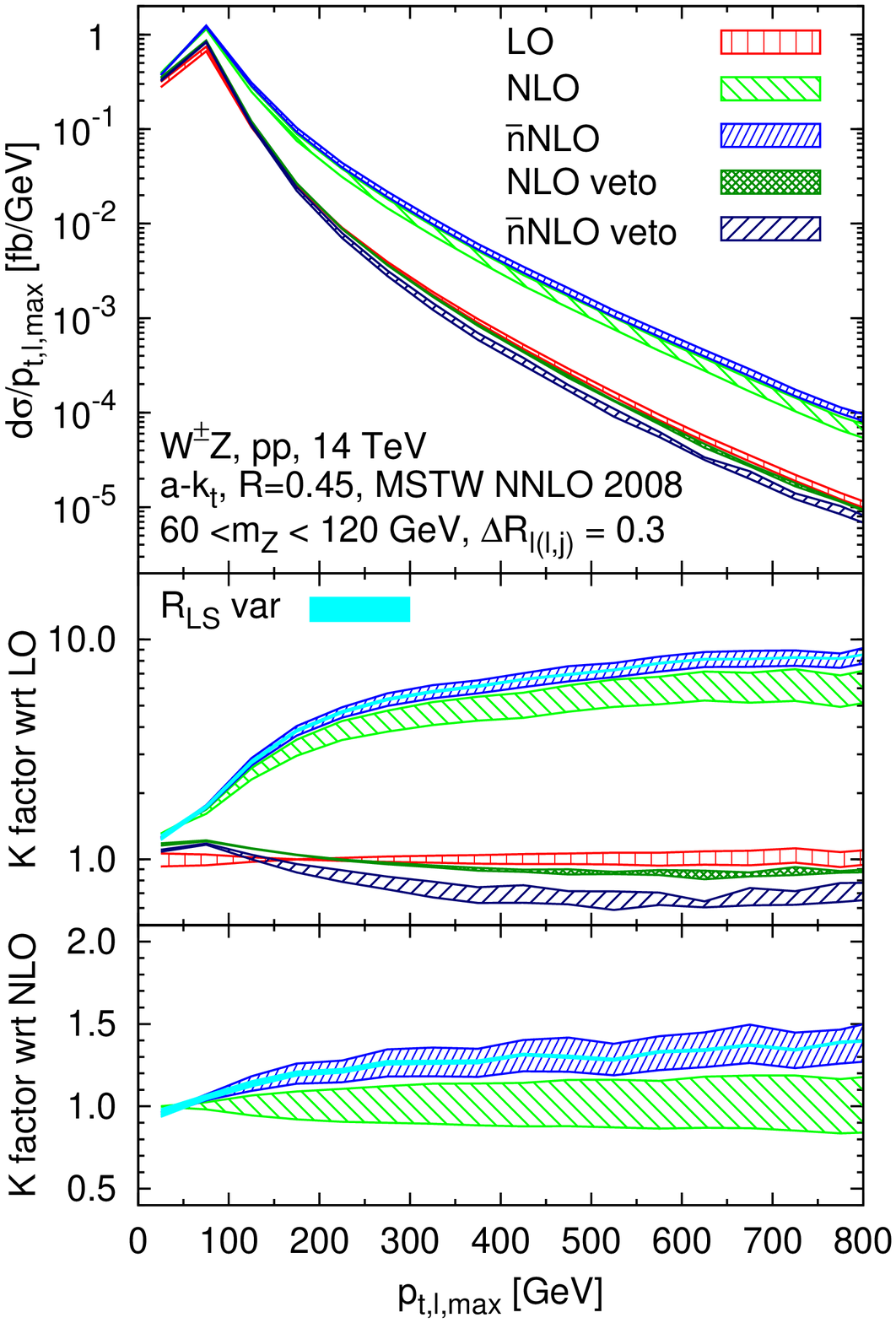}
  \caption{  
  Differential cross sections and K factors for the $p_T$ of the hardest lepton
  for the LHC at $\sqrt{s}=8\, \text{TeV}$ (left) and $\sqrt{s}=14\, \text{TeV}$
  (right). The bands correspond to varying $\mu_F=\mu_R$ by factors 1/2 and 2
  around the central value from Eq.~(\ref{eq:ren}). The cyan solid bands give
  the uncertainty related to the $R_\text{LS}$ parameter varied between 0.5 and
  1.5.  The distribution are sums of contributions from two unlike flavor decay
  channels, $ee\mu\nu_\mu$ and $\mu\mu e\nu_e$.
  }
  \label{fig:pt_max}
\end{figure*}

\begin{figure*} [t!]
  \centering
  \includegraphics[width=1.0\columnwidth]{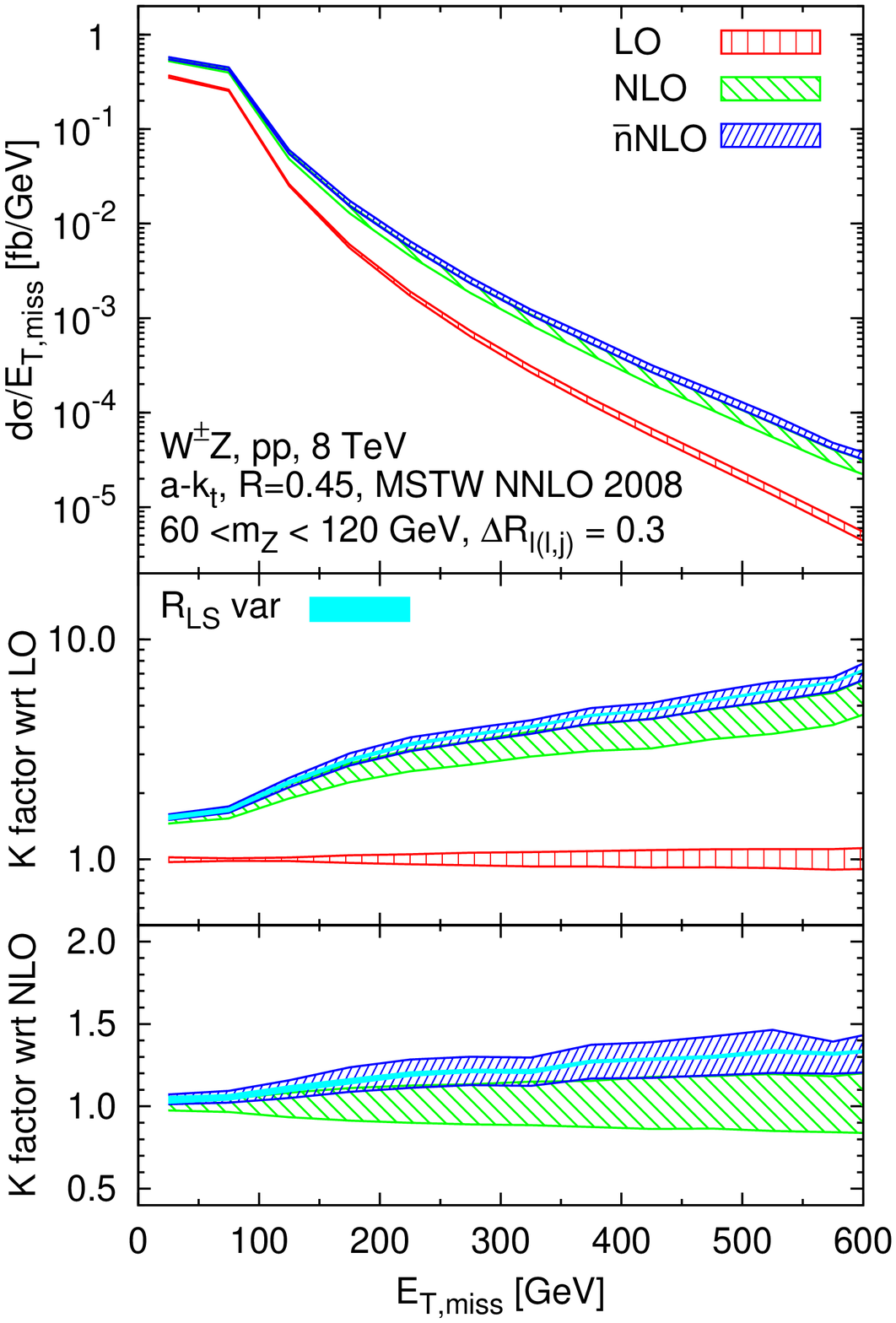}
  \hfill 
  \includegraphics[width=1.0\columnwidth]{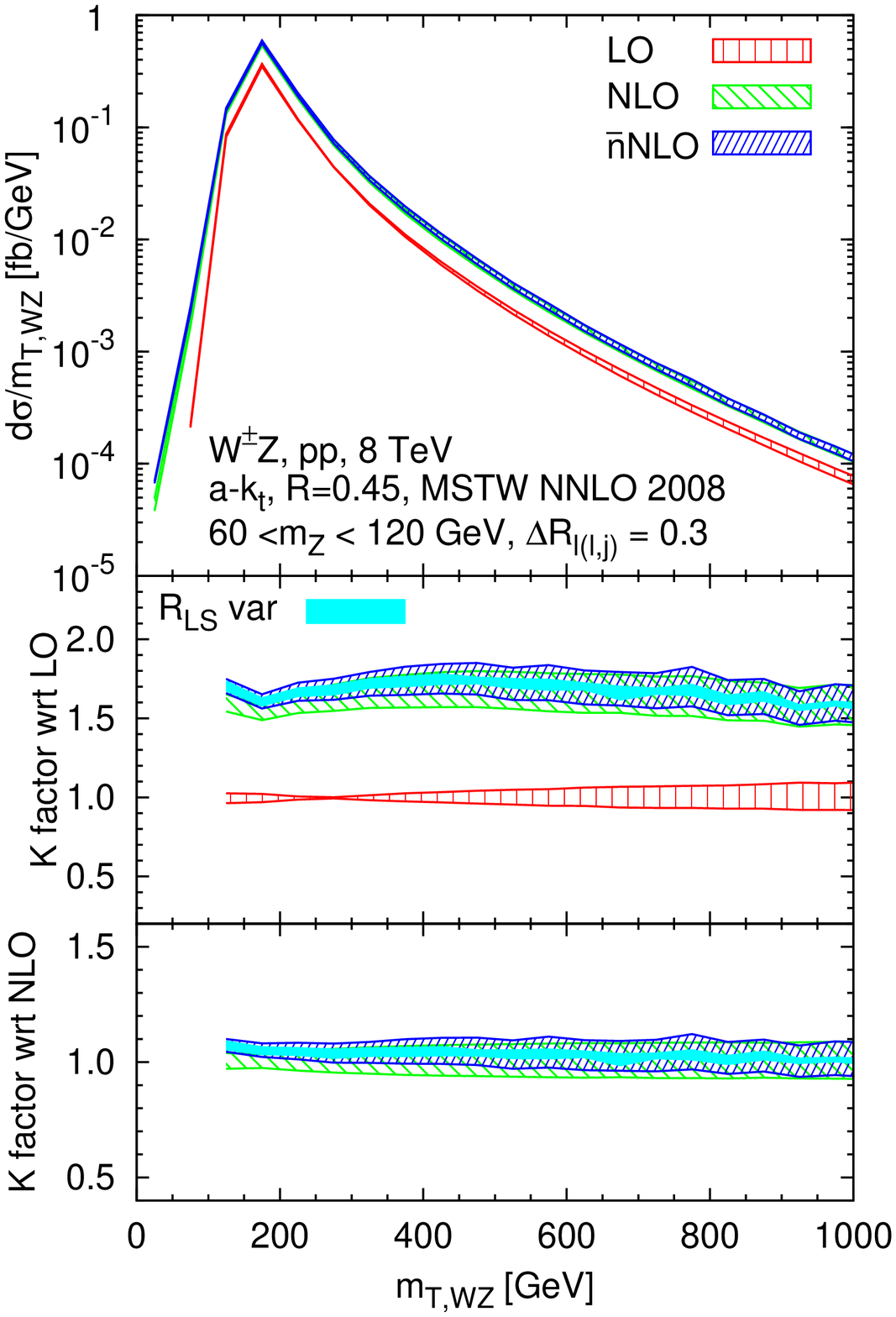}
  \caption{
  Differential cross sections and K factors for the missing transverse energy
  (left) and the transverse mass of the WZ system (right) for the LHC at 
  $\sqrt{s}=8\, \text{TeV}$. The bands correspond to varying $\mu_F=\mu_R$ 
  by factors 1/2 and 2 around the central value from Eq.~(\ref{eq:ren}).  The
  cyan solid bands give the uncertainty related to the $R_\text{LS}$ parameter
  varied between 0.5 and 1.5. The distribution are sums of contributions from
  two unlike flavor decay channels, $ee\mu\nu_\mu$ and $\mu\mu e\nu_e$.
  }
  \label{fig:ptnu}
\end{figure*}

In our computations, regardless of the order, we use the MSTW NNLO
2008~\cite{Martin:2009iq} PDFs with $\alpha_s(M_Z)= 0.11707$. 
We choose $m_Z=91.1876 \GeV$, $m_W=80.398 \GeV$ and
$G_F=1.16637\times 10^{-5}\GeV^{-2}$ as electroweak input parameters and
derive the weak mixing angle from the Standard Model tree level
relations. The center-of-mass energy is fixed to  $\sqrt{s} = 8 \TeV$ if not
specified otherwise.
As in the underlying VBFNLO WZ@NLO and WZj@NLO calculations, only the
channels with W and Z/$\gamma^*$ decaying into unlike flavor leptons are considered,
i.e. $ p p \rightarrow  W^\pm Z  +
X \rightarrow \ell_1^\pm
\stackrel{\text{\tiny(}-\text{\tiny)}}{\nu_1} \ell_2^+ \ell_2^-  + X$.
From the theoretical perspective, to take into account all possible decay
channels, i.e. $eee\nu_e$, $ee\mu\nu_\mu$, $\mu\mu e\nu_e$, $\mu\mu\mu\nu_\mu$,
one can safely multiply the single flavor result by a factor four, since the Fermi interferences due to
identical particles in the final state are below per mille
level~\cite{Campanario:2010hp}.  However, as explained later, we shall only
apply a cut on the reconstructed mass of the opposite charge same flavor
lepton pair.  Therefore, all the results shown in
Figs.~\ref{fig:HT}--\ref{fig:ptnu} correspond to the sum of contributions from
the two decay channels, $ee\mu\nu_\mu$ and $\mu\mu e\nu_e$.
All off-shell effects are included, which takes into account spin correlations
and also virtual photons. Finite width effects, related to leptonic decays of
the electroweak gauge bosons, are accounted for with a fixed-width
scheme~\cite{Campanario:2010hp}.
Top quark effects are not considered and all other quarks are taken massless. Effects
from generation mixing are neglected since the CKM matrix is set to the identity matrix. As
the central value for the factorization and renormalization scales we choose
\begin{equation}
\mu_{F,R}=
\frac12 \sum  p_{T,\text{partons}} + 
\frac12 \sqrt{p_{T,W}^2+m_W^2}+
\frac12 \sqrt{p_{T,Z}^2+m_Z^2}
\label{eq:ren}
\end{equation}
where $p_{T,V}$ and $m_V$ have to be understood as the reconstructed
transverse momenta and invariant masses of the decaying bosons.
To study the impact of the QCD corrections, we choose inclusive cuts.  The
charged leptons are required to be hard and central: $p_{T,\ell}\ge 15(20)$, for
$l$ coming from Z(W), and $|y_{l}|\le 2.5$. The missing transverse energy must
satisfy the cut $E_{T,\text{miss}} > 30 \GeV$.  The reconstructed mass from the
opposite-charge same-flavor leptons has to lie in the window $60 < m_{l^+l^-} <
120 \GeV$, which also avoids singularities coming from off-shell photons,
$\gamma^*\to \ell^+ \ell^-$. We cluster all final state partons with $|y_p|\le 5
$ to jets with the anti-$k_t$ algorithm~\cite{Cacciari:2008gp}, as implemented
in FastJet~\cite{Cacciari:2005hq, FastJet}, with the radius $R=0.45$.
For observables that involve jets, we
  consider only those jets that lie in the rapidity range
  $|y_{\text{jet}}|\le 4.5$ and have transverse momenta 
  $p_{T,\text{jet}}\ge 30\GeV$.
In addition, we impose a requirement on the lepton-lepton and lepton-jet
separation in the azimuthal angle-rapidity  plane $ \Delta R_{l(l,j)} > 0.3$.

As a first check of our setup, we have merged WZ@LO and WZj@LO to
produce WZ@\nLO, which can be tested against the full WZ@NLO result. 
In Fig.~\ref{fig:HT}, we show the effective mass defined by
\begin{equation}
H_{T} = \sum  p_{T,\text{jets}} + \sum  p_{T,l} +
E_{T,\text{miss}}\,,
\label{eq:HT}
\end{equation}
which often enters in super-symmetry searches~\cite{CMS-ss,Aad:2011vj}.
This distribution is very sensitive to the enhancements from soft or collinear
emissions of the electroweak bosons as well as to additional parton radiation
coming from new channels.

In the middle panel of Fig.~\ref{fig:HT}, we show the K factors with
respect to the LO result and, in the bottom, the corresponding ratios to NLO.
At low $H_T$ values, the difference between \nLO and NLO is at the level of
30$-40\%$. This is due to the fact that  we do not have control over the finite
virtual constant terms which are missing in our \nLO approximation. 
However, as the $H_T$ increases, the \nLO result converges to the full NLO,
providing a prediction more accurate than a simple LO calculation. Note in the
middle panel the large $\nLO$ corrections and fast convergence to the NLO
result, yielding K-factors of order of 10. 
The $\nNLO$ corrections can be as large as 100$\%$ compared to NLO (bottom panel
of Fig.~\ref{fig:HT}) and they are clearly beyond the NLO scale uncertainties.
We observe that the R$_\text{LS}$ uncertainties are small and there is only a
marginal reduction in the scale uncertainties at  \nNLO.  The latter is due to
the fact that we are favoring regions of the phase space associated with new
topologies entering at NNLO which are only computed at LO.  

The integrated cross section for one lepton flavor, $\sigma(pp \to l_1 \nu_{l_1}
l_2 l_2)$, dominated by the low $H_T$ region, increases only by about  5$\%$
from 25.7 $\pm$ 1.0 (scale) fb at NLO to  26.9 $\pm$ 0.9 (scale) $\pm $ 0.4
(R$_\text{LS}$)~fb at \nNLO.
We have also computed the total cross sections for 7~TeV at NLO and
the result agrees with this quoted in \cite{CMS-WZ}.

In Fig.~\ref{fig:pt_max}, we show differential distributions for the lepton with
maximum $p_T$ for $\sqrt{s}=$ 8 and $\sqrt{s}=$ 14 $\TeV$, left and right panels
respectively. We observe that the $\nNLO$ corrections are large at both
energies. For the 14 $\TeV$ run, these corrections are beyond the
renormalization and factorization scale uncertainties above $p_T=$ 150 GeV. 
The same is true for the 8 $\TeV$ case only at slightly higher $p_T$. 
Even though the effect is less pronounced than at 14 TeV, due to
  the smaller relative importance of additional parton radiation, the corrections are at the 15$\%$ level
already at 200 $\GeV$. 
In addition, for the 14~TeV case, we show the curves with a veto on the jets at
NLO and $\nNLO$ where we require the absence of any jets with $p_T > 50$~GeV.
The $\nNLO$-veto corrections are negative and beyond the scale uncertainties of
the NLO-vetoed predictions. This might be of relevance for anomalous coupling
searches since vetoed distributions are often used to suppress additional
radiation.  The entire $\nNLO$-veto result has larger scale uncertainties than
the NLO-vetoed curves, thus revealing, partially, accidental cancellation
happening at NLO.
Such a feature has been extensively discussed for jet vetoes also in the context
of NNLO calculations of Higgs-boson
production, e.g.~\cite{Banfi:2012yh} and references therein.

In Fig.~\ref{fig:ptnu} (left), we show the differential distributions for the
missing transverse energy. We see that the $\nNLO$ corrections to this
observable can be as
large as the 30$\%$, exceeding the scale uncertainty of the NLO result. 
In the right-hand plot of Fig.~\ref{fig:ptnu}, we
show the cluster energy (transverse mass of the WZ system) defined by
\begin{equation}
m_{T, \text{WZ}} =
\sqrt{(E_{T}^W+E_{T}^Z)^2-(p_{x}^W+p_{x}^Z)^2-(p_{y}^W+p_{y}^Z)^2}\,,
\end{equation}
where $E^{W,Z}_T$ and $p^{W,Z}_{x,y}$ are the transverse energy and the
transverse momentum components of the bosons reconstructed from the four-momenta
of the leptons.
The \nNLO corrections to this observable are small, which means that this
distribution is not particularly sensitive to the new topologies that appear at
NNLO.  For the same reason, the finite terms from the two-loop diagrams, which
are missing in the \nNLO result, are of larger relative importance for this
observable than they are for $H_T$, $p_{T,l,\max}$ or $E_{T,\text{miss}}$, which
is also reflected in larger $R_\text{LS}$ uncertainties in the $m_{T,
\text{WZ}}$ case.

The reason behind the marked difference in the relative size of the \nNLO
corrections between $m_{T, \text{WZ}}$ and all the other observables which we
have discussed is the following. For observables like $p_{T,l,\max}$ or
$E_{T,\text{miss}}$, to receive large contributions in the high-$p_T$ region, it
is enough that only one of the two bosons is produced at high $p_T$. This
high-$p_T$ boson recoils against a high-$p_T$ QCD parton. 
The other boson can be in particular soft and collinear to a quark and such
configurations, one of which is depicted in the bottom diagram of
Fig.~\ref{fig:diagram}, come with the $\as^2\aEW^2\ln^2\left(p_{T,j}/m_V\right)$
enhancement. In the case of $m_{T, \text{WZ}}$, however, the favored
configurations are those in which both bosons have sizable $p_T$s and are
preferably back to back therefore do not lead to logarithmic enhancements at
NNLO.

%------------------------------------------------------------------------------
\section{Conclusions}
\label{sec:concl}

In this letter, we have used LoopSim together with VBFNLO to compute an
approximation to the NNLO QCD corrections for the process $pp \rightarrow
\ell_1^\pm \smash{\stackrel{\text{\tiny(}-\text{\tiny)}}{\nu_1}} \ell_2^+
\ell_2^- + X$. 
Our result, referred to as \nNLO, is expected to be accurate in the regions of
phase space dominated by the topologies other than those present at LO.
As in the underlying VBFNLO WZ@NLO and WZj@NLO calculations, our
treatment includes the leptonic decays of the vector bosons and all off-shell and finite-width effects.

We found that the $\nNLO$ corrections to a number of observables are sizable
at high $p_T$ and have non-trivial kinematic dependence.
It is therefore important to take them into account in searches for physics
beyond the SM and other physics analyses that involve  WZ production. 
The VBFNLO+LoopSim code is available from the authors on request and it will be
made public in the near future.

%%%%%%%%%%%%%%%%%%%%%%%%%%
\section*{Acknowledgments} 
We thank Gavin Salam for collaboration during the initial stages of this
work and for subsequent discussions and comments on the manuscript.
FC acknowledges support by FEDER and Spanish MICINN under grant FPA2008-02878
and by the Deutsche Forschungsgemeinschaft under SFB TR-9 ``Computergest\"utzte
Theoretische Teilchenphysik''.
SS was in part supported by European Commission under contract
PITN-GA-2010-264564.

%%%%%%%%%%%%%%%%%%%%%%%%%%%%%%%%%%%%%%%%%%%%%%%%%%%%%%%%%%%%%%

\end{document}